\documentclass[twoside,10pt,journal,compsoc]{IEEEtran}

\ifCLASSOPTIONcompsoc
  \usepackage[nocompress]{cite}
\else
  \usepackage{cite}
\fi
\usepackage{amsmath}
\usepackage{amssymb}  
\usepackage{amsthm}
\usepackage{algpseudocode}
\usepackage{algorithm}
\usepackage{array}
\usepackage{tcolorbox}
\usepackage{url}
\usepackage{graphics} 
\usepackage{epsfig} 
\usepackage{enumerate}
\usepackage{booktabs}
\usepackage[bottom]{footmisc}
\usepackage{enumitem}
\setlist{nosep}
\setlist[itemize]{leftmargin=*}

\usepackage[font=small,labelfont=bf]{caption}

\usepackage{hyperref}
\hypersetup{
	colorlinks=true, 
    linkcolor=blue,  
    citecolor=blue,  
    filecolor=blue,  
    urlcolor=blue    
}

\DeclareMathOperator*{\argmax}{arg\,max}

\newcommand{\AoI}{\Delta}
\newcommand{\sysTime}{H}

\begin{document}

\title{WiFresh: Age-of-Information from\\ Theory to Implementation}

\author{Igor Kadota, Muhammad Shahir Rahman, and Eytan Modiano
\IEEEcompsocitemizethanks{\IEEEcompsocthanksitem Igor Kadota is with the Department
of Electrical Engineering, Columbia University, New York, NY, 10027.\protect\\
E-mail: igor.kadota@columbia.edu
\IEEEcompsocthanksitem Muhammad Shahir Rahman and Eytan Modiano are with the Laboratory for Information and Decision Systems (LIDS), Massachusetts Institute of Technology, Cambridge, MA, 02139.\protect\\
E-mail: shahir@mit.edu and modiano@mit.edu}
\thanks{This paper was presented in part at ACM MobiCom 2020 \cite{igorWiFreshPoster}.}
}



\IEEEtitleabstractindextext{%
\begin{abstract}
Emerging applications, such as smart factories and fleets of drones, increasingly rely on sharing time-sensitive information for monitoring and control. In such application domains, it is essential to keep information fresh, as outdated information loses its value and can lead to system failures and safety risks. The Age-of-Information is a performance metric that captures how fresh the information is from the perspective of the destination. 

In this paper, we show that as the congestion in the wireless network increases, the Age-of-Information degrades sharply, leading to outdated information at the destination. Leveraging years of theoretical research, we propose WiFresh: an unconventional architecture that achieves near optimal information freshness in wireless networks of any size, even when the network is overloaded. Our experimental results show that WiFresh can improve information freshness by two orders of magnitude when compared to an equivalent standard WiFi network. We propose and realize two strategies for implementing WiFresh: one at the MAC layer using hardware-level programming and another at the Application layer using Python.
\end{abstract}

}

\maketitle

\IEEEdisplaynontitleabstractindextext

\IEEEpeerreviewmaketitle

\section{Introduction}\label{sec5.Intro}
Emerging applications will increasingly rely on sharing time-sensitive information for monitoring and control. Examples are abundant: 
monitoring mobile ground-robots in automated fulfillment warehouses at Amazon \cite{Kiva,KivaWiFi} and Alibaba \cite{Alibaba}; 
collision prevention applications \cite{DSRC} for vehicles on the road \cite{PerformanceDSRC,SimVsRealDSRC,VehicleToInfrastructureDSRC,AppLevelDSRC}; 
path planning, localization and motion control for multi-robot formations using drones \cite{Drones,DistributedRobotFormation} and using ground-robots \cite{RobotFormation}; 
multi-drone system for tracking a mobile spectrum cheater \cite{ASTRO}; 
multi-drone system for automated aerial cinematography \cite{DroneCinema}; 
multi-drone system for exploration of subterranean environments \cite{DARPASubTDrones};
multi-robot simultaneous localization and mapping (SLAM) using drones \cite{DOOR-SLAM,CooperativeSLAM} and using ground-robots \cite{MultiRobotSlam}; 
real-time surveillance system using a fleet of ground-robots \cite{MultiRobotMapping}; and
data collection from sensors, drones and cameras for agriculture using the Azure FarmBeats IoT platform \cite{FarmBeats,FarmBeats2}. 
In such application domains, it is essential to \emph{keep information fresh}, as outdated information loses its value and can lead to \emph{system failures} and \emph{safety risks}. 

\emph{The various time-sensitive applications in}  \cite{Kiva,KivaWiFi,Alibaba,PerformanceDSRC,SimVsRealDSRC,VehicleToInfrastructureDSRC,AppLevelDSRC,Drones,DistributedRobotFormation,RobotFormation,ASTRO,DroneCinema,DOOR-SLAM,CooperativeSLAM,MultiRobotSlam,DARPASubTDrones,MultiRobotMapping,FarmBeats,FarmBeats2} \emph{are all implemented using the IEEE 802.11 standard (WiFi)}. WiFi is an attractive choice for it is low-cost, well-established, and immediately available in drones \cite{DistributedRobotFormation,DroneCinema}, computing platforms running the Robot Operating System (ROS) \cite{MultiRobotMapping}, sensors that measure soil temperature, pH, and moisture \cite{FarmBeats,FarmBeats2}, and in the Raspberry Pis used in this paper. Moreover, as showcased by these various implementations, \emph{small-scale underloaded} WiFi networks are capable of supporting time-sensitive applications. Two main shortcomings of WiFi, or any other wireless technology employing First-Come First-Served (FCFS) queues and Random Multiple Access mechanisms, are scalability and congestion.

\begin{figure}[t]
\begin{center}
\resizebox{.7\columnwidth}{!}{\includegraphics{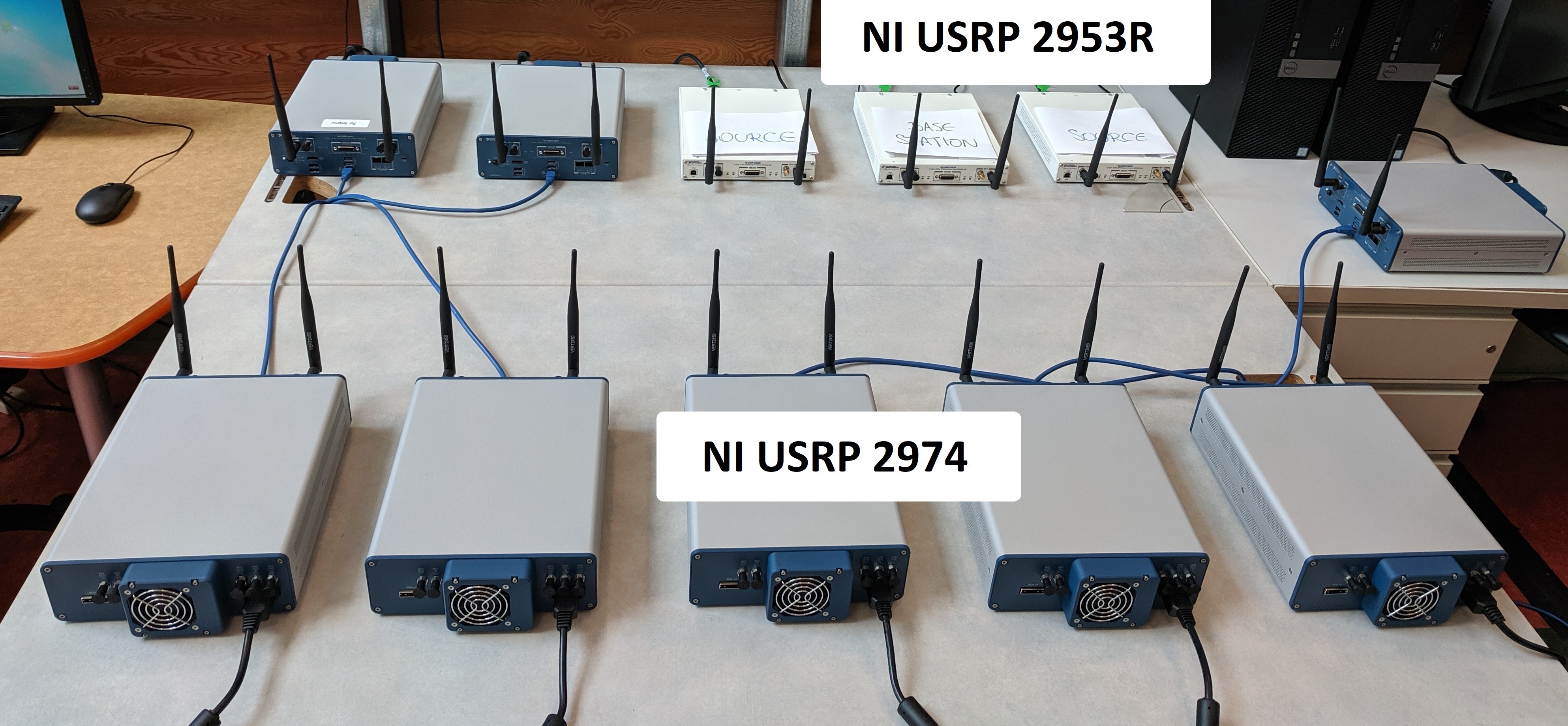}}
\end{center}
\vspace{-0.2cm}
\caption{\label{fig5.WiFreshRTtestbed} WiFresh RT testbed.}
\vspace{-0.2cm}
\end{figure}

\begin{figure}[t]
\begin{center}
\resizebox{.55\columnwidth}{!}{\includegraphics{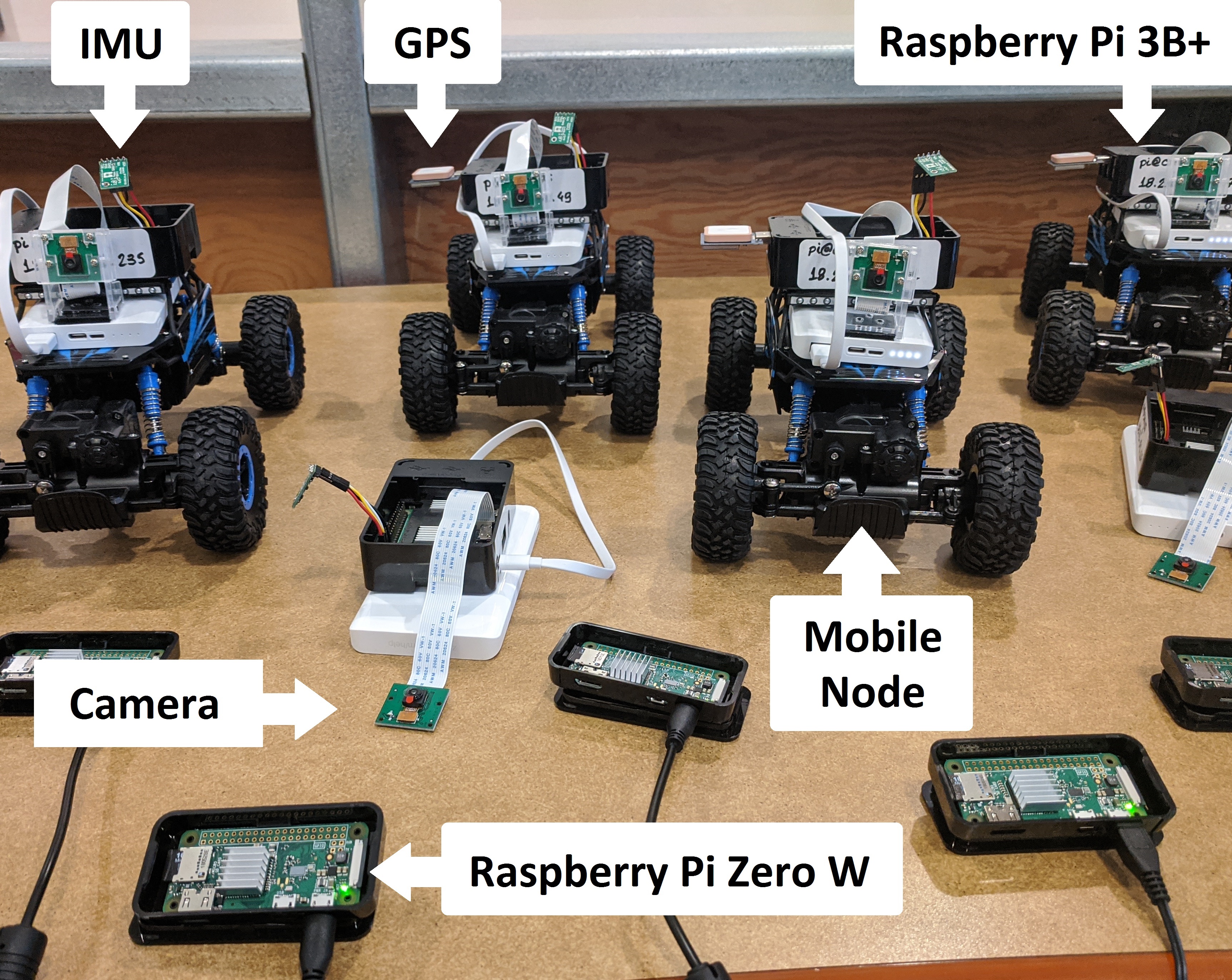}}
\end{center}
\vspace{-0.2cm}
\caption{\label{fig5.WiFreshAppTestbed} {WiFresh App sources and sensors.}}
\vspace{-0.2cm}
\end{figure}

\textbf{Our contribution}: 1) Leveraging years of theoretical research on Age-of-Information, we propose WiFresh: an unconventional network architecture that scales gracefully, achieving near optimal information freshness in wireless networks of any size, even when the network is overloaded.
2) We propose and realize two strategies for implementing WiFresh: \underline{WiFresh Real-Time}, which is designed to maximize performance, and is implemented at the \emph{MAC layer} in a network of eleven FPGA-based Software Defined Radios (Fig.~\ref{fig5.WiFreshRTtestbed}) using hardware-level programming; and \underline{WiFresh App}, which is designed to lower the barriers to adoption, and is implemented at the \emph{Application layer}, without modifications to lower layers of the networking protocol stack, in a network of twenty five Raspberry Pis (Fig.~\ref{fig5.WiFreshAppTestbed}) using Python~$3$. The WiFresh App runs over UDP and standard WiFi, making it easy to integrate into applications that are implemented using WiFi such as \cite{Kiva,KivaWiFi,Alibaba,PerformanceDSRC,SimVsRealDSRC,VehicleToInfrastructureDSRC,AppLevelDSRC,Drones,DistributedRobotFormation,RobotFormation,ASTRO,DroneCinema,DOOR-SLAM,CooperativeSLAM,MultiRobotSlam,DARPASubTDrones,MultiRobotMapping,FarmBeats,FarmBeats2}. 
3) Our experimental results in \textsection\ref{sec5.evaluation} show that the more congested the network, the more prominent is the superiority of WiFresh when compared to WiFi. In particular, we show that under high load, WiFresh can improve information freshness by two orders of magnitude when compared to an equivalent standard WiFi network. 

\begin{figure}[t]
\begin{center}
\resizebox{.99\columnwidth}{!}{\includegraphics{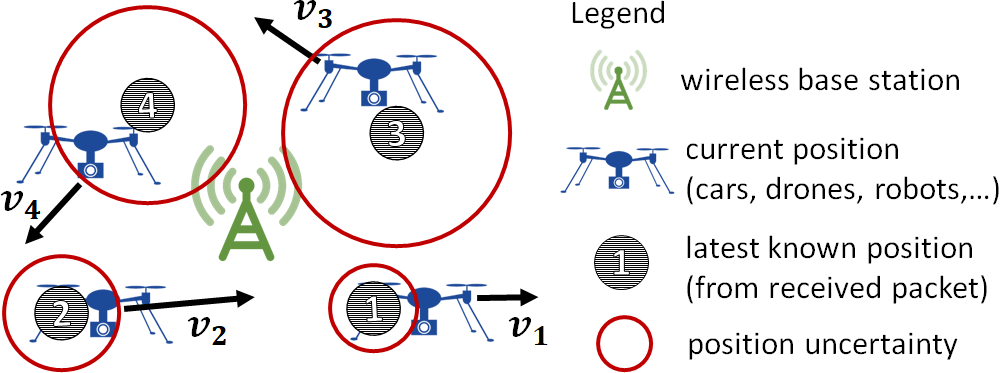}}
\end{center}
\vspace{-0.2cm}
\caption{\label{fig5.illustrative_network} Illustration of a monitoring system.} 
\vspace{-0.2cm}
\end{figure}

To illustrate the concept of \emph{information freshness}, which is formally defined in \textsection\ref{sec5.background},
consider a monitoring system composed of a remote monitor, a wireless base station (BS) and $N$ mobile nodes. Each node $i\in\{1,2,\cdots,N\}$ moves with an average velocity of $v_i$ meters per second, generates status information from time to time, and sends this information to the remote monitor via the wireless base station. Status information can include the node's current position, inertial measurements, and pictures of the environment. The remote monitor keeps track of the information, and is particularly interested in the position of the nodes. Assume that at time~$t$, the latest packet received by the remote monitor from node $i$ had information about its position at time $\tau_i(t)$. Then, the quantity $\AoI_i(t):=t-\tau_i(t)$ captures how fresh the position information is at time $t$. We refer to $\AoI_i(t)$ as \emph{Age-of-Information (AoI)} and say that the remote monitor has \emph{stale information} when $\AoI_i(t)$ is large, and \emph{fresh information} when $\AoI_i(t)$ is small. In particular, an age of $\AoI_i(t)=2$ seconds represents that at time $t$ the remote monitor knows the location of node $i$ two seconds ago. Hence, the \emph{uncertainty about node} $i$\emph{'s position at time} $t$ is captured by the quantity $v_i\AoI_i(t)$, as illustrated in Fig.~\ref{fig5.illustrative_network}, and a \emph{large age corresponds to a large uncertainty}.

In Table~\ref{tab5.position_uncertainty_WiFresh}, we consider a sequence of networks with increasing size $N$ and display the average age $\AoI_i(t)$ in seconds, where the average is taken over time $t$ and over all the $N$ sources. Each \emph{source} is a Raspberry Pi (shown in Fig.~\ref{fig5.WiFreshAppTestbed}) generating position information using the Stratus GPYes 2.0 u-blox~8 GPS receiver and generating inertial measurements using the Pololu MinIMU-9 v5 sensor. The \emph{wireless base station} is a Raspberry Pi receiving data from the $N$ sources. We compare networks using three different architectures, namely:
\begin{enumerate}
\item WiFi UDP, which is UDP over standard WiFi; 
\item WiFi Age Control Protocol (ACP), which uses the Transport layer protocol developed in \cite{AoI_ACP} to \emph{control the packet generation rates} at the sources in order to minimize AoI;
\item WiFresh App, which is one of the WiFresh implementations proposed in this paper.
\end{enumerate}
Additional details about the experimental setup are provided in \textsection\ref{sec5.Network_Scalability}.

\begin{table}[t]
\caption{\label{tab5.position_uncertainty_WiFresh} Measurements of the average age $\AoI_i(t)$ (in seconds) for networks with different number of sources $N$.} 
\vspace{-0.2cm}
\begin{center}
\begin{tabular}{cccccc} \toprule
$N$ & 2 & 8 & 12 & 20 & 24 \\
\midrule
WiFi UDP& $0.34$ & $0.32$ & $0.32$ & $22.43$ & $34.09$ \\
WiFi ACP& $0.99$ & $0.97$ & $0.88$ & $5.76$  & $6.91$  \\
WiFresh App & $0.29$ & $0.35$ & $0.39$ & $0.49$  & $0.54$  \\
\bottomrule
\end{tabular}
\end{center}
\vspace{-0.2cm}
\end{table}

\textbf{WiFi UDP.} The measurements in the \emph{first row} of Table~\ref{tab5.position_uncertainty_WiFresh} show that as the number of sources $N$ in the WiFi UDP network increases, the \emph{network becomes overloaded and the average age} $\AoI_i(t)$ \emph{degrades sharply}. The average age for $N=12$ sources is $0.32$ seconds, while for $N=20$ sources is $22.43$ seconds, which means that the information at the remote monitor is (on average) $22$ seconds old. Naturally, this staleness directly affects the capability of the monitoring system of tracking the current position of the source nodes, especially for source nodes that are moving fast.

\textbf{WiFi ACP.} The \emph{second row} of Table~\ref{tab5.position_uncertainty_WiFresh} shows that the average age $\AoI_i(t)$ for WiFi ACP with $N=20$ sources is $5.76$ seconds. By controlling the packet generation rates at the sources using the protocol developed in \cite{AoI_ACP}, ACP \emph{improves the average age by a factor of four} when compared with WiFi UDP. Notice that a high packet generation rate may overload the network and lead to a high average age, while a low packet generation rate may result in infrequent information updates at the destination, which may also lead to a high average age. The ACP dynamically adapts the packet generation rates at the sources in order to drive the network to the point of optimal information freshness. This point of minimum AoI is illustrated in Fig.~\ref{fig5.AoI_tradeoff_plot}. 

\textbf{WiFresh App.} The \emph{third row} of Table~\ref{tab5.position_uncertainty_WiFresh} shows that the average age $\AoI_i(t)$ for WiFresh with $N=20$ sources is $0.54$ seconds. WiFresh \emph{improves the average age by a factor of forty} when compared with WiFi UDP. Experimental results in \textsection\ref{sec5.evaluation} show that this improvement increases for larger $N$.
The superior performance of WiFresh is due to the combination of three elements: 
\begin{itemize}
\item \textbf{Polling Multiple Access mechanism} that \emph{prevents packet collisions}, allowing for efficient resource allocation among sources, which is critical in congested networks and in networks with large number of sources $N$; 
\item \textbf{Max-Weight (MW) policy} that determines the sequence of sources to poll in order to keep the age of each source as low as possible and, thus, \emph{optimize information freshness} in the network. Notice that the Polling mechanism is used to support the MW policy; and 
\item \textbf{Last-Come First-Served (LCFS) queues} that prioritize the \emph{packet with lowest delay}, leading to sources that always transmit the freshest packets to the base station. 
\end{itemize}
In WiFresh App, these three elements are implemented at the Application layer in a network of Raspberry Pis. In WiFresh Real-Time, these three elements are implemented at the MAC layer in a network of SDRs. 
The choice of each of these elements is underpinned by theoretical research. In \cite{AoI_management,AoI_LIFO,AoI_LGFS19_2}, the LCFS queue was shown to be the optimal queueing discipline in terms of AoI in different settings. In \cite{AoI_sync,igorMobiHoc,igorTON18,Book19}, the authors developed  performance guarantees for the MW policy in different settings.


\textbf{Scalability problem.} Neither WiFi UDP nor WiFresh attempt to control the packet generation rate at the sources. Hence, when the number of sources $N$ increases to the point that the cumulative packet generation rate exceeds the capacity of the network, both WiFi UDP and WiFresh become overloaded and the number of backlogged packets at the sources grows rapidly. For WiFi UDP, a large backlog in the First-Come First-Served (FCFS) queues leads to high packet delay and, thus, to high average age, as observed in Table~\ref{tab5.position_uncertainty_WiFresh}. In contrast, \emph{WiFresh scales gracefully, even when the network is overloaded}, with average age increasing linearly with $N$. In \cite[Chapter~3]{Book19}, the authors derived a lower bound on the achievable AoI performance in wireless networks and concluded that the average age cannot scale better than linearly on the network size $N$.
%

The remainder of this paper is organized as follows. In \textsection\ref{sec5.related}, we describe related work on AoI. In \textsection\ref{sec5.background}, we discuss the impact of the multiple access mechanism, transmission scheduling policy, and queueing discipline on information freshness. In \textsection\ref{sec5.design}, we describe the design and implementation of WiFresh Real-Time and WiFresh App. In \textsection\ref{sec5.evaluation}, we evaluate the performance of WiFresh in networks with increasing load and increasing size. The paper is concluded in \textsection\ref{sec5.conclusion}.

\section{Related Work}\label{sec5.related}

The Age-of-Information was recently proposed in \cite{AoI_update} and has been receiving increasing attention in the literature for its application in communication systems that carry time-sensitive data. The AoI captures how fresh the information is \emph{from the perspective of the destination}, in contrast to the well-established packet delay that represents the latency of \emph{a particular packet}. Most papers on AoI focus on theory and a few consider system implementation.

\textbf{Theoretical research.} The problem of optimizing AoI has been addressed in a variety of contexts. Queueing Theory is used in \cite{AoI_update,AoI_MG1,AoI_management,AoI_path,AoI_nonlinear} to characterize the AoI performance of various important queueing systems. Game Theory is used in \cite{AoI_game_interference,AoI_game_coexistence,AoI_game_jamming} to analyze the impact of communication interference on AoI. Information Theory is used in \cite{AoI_IT_Yates,AoI_IT_best} for designing source and channel coding schemes to improve AoI. Optimization of scheduling policies 
in communication networks is considered in \cite{igorMobiHoc,AoI_scheduling,igorTON18_2,Rajat17,YuPin17,igorTON18,AoI_multiaccess,AoI_discipline,YinSun18}. 
Different applications of AoI in sensor networks, cellular networks and vehicular networks are analyzed and/or emulated in \cite{AoI_emulation,AoI_young,AoI_LUPMAC,BaiochiAoI}. This list of works is not exhaustive. For a more comprehensive list of theoretical works we refer the reader to \cite{Yin_website,Book19,Book_AoI_17}.
 
The theoretical work that is most relevant to this paper is our prior work in \cite{igorMobiHoc} which considers a downlink wireless network with a base station generating packets according to a stochastic process and transmitting these packets to different destinations. The analysis in \cite{igorMobiHoc} shows that the performance of a network employing LCFS queues and MW policy is guaranteed to be within a factor of four away from the minimum achievable AoI. The main differences between \cite{igorMobiHoc} and this paper are with respect to the network setting, the assumptions, and the objective. This paper considers an uplink wireless network in which multiple sources generate data packets and simultaneous transmissions can cause packet collisions, introducing the need for a multiple access mechanism. This paper assumes that information sources are not synchronized, that the packet generation process at the sources is unknown, and that the channel quality of the links is dynamic and unknown. Moreover, the goal of this paper is to bridge theory and practice. In particular, we leverage the theoretical results in the AoI literature to propose and implement a \emph{practical} architecture that can keep the information in the network as fresh as possible.  
 
\textbf{Systems research.} A few papers \cite{AoI_measurements,AoI_VANET,AoI_ACP,igorWiFreshPoster} have implemented AoI-based systems. In \cite{AoI_measurements}, the authors consider a source-destination pair transmitting packets over the Internet and measure the AoI for different packet generation rates. In \cite{AoI_VANET}, the authors consider a vehicular network and develop an Application layer algorithm that adapts the packet generation rates at the sources to improve information freshness. This algorithm is validated using the ORBIT testbed with wireless sources employing WiFi, in particular the IEEE 802.11a standard. In \cite{AoI_ACP}, the authors consider an Internet-of-Things network and develop a Transport layer protocol named Age Control Protocol (ACP) that adapts the packet generation rates at the sources in order to optimize for information freshness. This protocol is validated using ten sources connected via WiFi to the Internet and sending packets to a destination in another continent. In \textsection\ref{sec5.evaluation}, we implement ACP and evaluate its performance against WiFresh. Notice that  \cite{AoI_measurements,AoI_VANET,AoI_ACP} address the problem of controlling the packet generation rates at the sources in order to optimize information freshness.

In this paper, we develop and implement a network architecture that is optimized across the queueing discipline, the multiple access mechanism, and the transmission scheduling policy. 
A simplified version of WiFresh Real-Time was introduced in our poster \cite{igorWiFreshPoster}.
The main differences between \cite{igorWiFreshPoster} and this paper are with respect to the scope. This paper presents an in-depth description of WiFresh Real-Time and WiFresh App, with discussions about the impact of the multiple access mechanism, transmission scheduling policy, and queueing discipline on AoI. This paper addresses issues pertaining to packet fragmentation and to sources that generate multiple information types such as position, inertial measurements, and images. This paper provides extensive experimental results for networks with different sizes and different traffic loads, and compares the performance of WiFresh with WiFi UDP, WiFi TCP, WiFi ACP, and others.



\section{Background on AoI}\label{sec5.background}
\begin{figure}[t]
\begin{center}
\resizebox{.8\columnwidth}{!}{\includegraphics{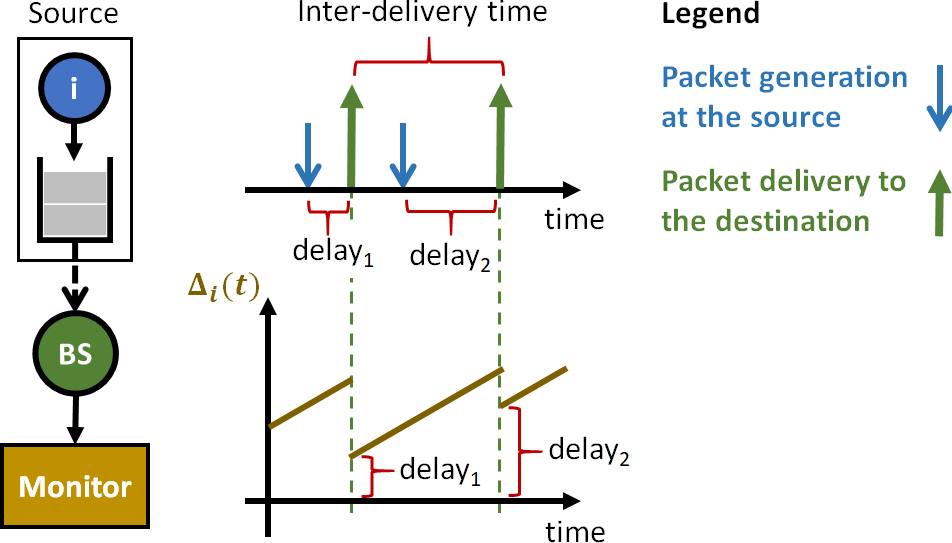}}
\end{center}
\vspace{-0.2cm}
\caption{\label{fig5.AoI_illustration} AoI evolution for a single information source sending packets to the remote monitor via the BS.} 
\vspace{-0.3cm}
\end{figure}

Consider a communication network in which packets are time-stamped upon arrival. Naturally, the higher the time-stamp, the fresher is the information contained in a packet. Let $\tau_i(t)$ be the time-stamp of the \emph{freshest packet received by the destination from source $i$ by time $t$}. 
%
The AoI associated with source $i$ is defined as $\AoI_i(t):=t-\tau_i(t)$. The AoI measures the time elapsed since the generation of the freshest packet received by the destination from source $i$.
The value of $\AoI_i(t)$ increases linearly in time while no fresher packet from source $i$ is received, representing the information getting older. At the moment a \emph{fresher} packet is received by the destination, the value of $\tau_i(t)$ is updated and $\AoI_i(t)$ \emph{decreases to the packet delay}. The evolution of the age process is illustrated in Fig.~\ref{fig5.AoI_illustration}. 

The time-average expected age associated with source~$i$ is given by $\int_{t=0}^T\mathbb{E}[\AoI_i(t)]dt/T$. 
From Fig.~\ref{fig5.AoI_illustration}, we can see that to keep the information at the destination as fresh as possible, i.e. minimize the time-average expected AoI in the network, it is necessary to simultaneously provide: i) low packet delay; ii) high data throughput; and iii) service regularity\footnote{It is important to emphasize the difference between delivering packets regularly and providing a minimum throughput. In general, a given minimum throughput can be achieved even if long periods with no delivery occur, as long as those are balanced by short periods of consecutive packet deliveries.}. 
To minimize AoI, we consider the network as a whole and optimize the system across the queueing discipline, the multiple access mechanism, and the transmission scheduling policy. Next, we discuss each of them in detail.


\subsection{Queueing Discipline}\label{sec5.Queueing}
The queueing discipline employed at the sources is central for minimizing AoI. In this section, we compare FCFS and LCFS queues and evaluate their performance in terms of AoI. FCFS queues are widely deployed in communication systems and they are the basis for other disciplines such as Priority Queueing and Fair Queueing. FCFS queues transmit packets in order of arrival, meaning that the \emph{freshest packet is always placed at the tail of the queue}. 
Under heavy loads, the FCFS queue is often backlogged and the freshest packet has to wait for a long queueing delay before being delivered to the destination. The high queueing delay leads to stale information at the destination and to high age. This effect is more prominent for large FCFS queues, as discussed in \textsection\ref{sec5.Single_Source}.

LCFS queues are often considered in the AoI literature  \cite{AoI_discipline,AoI_management,AoI_LIFO,AoI_LGFS19_1,AoI_LGFS19_2,igorMobiHoc}, but they are not commonly deployed in communication systems. LCFS queues place the \emph{most recently generated packet at the Head-of-Line (HoL)}, leading to sources that transmit the freshest packet first, which makes LCFS queues ideal for applications that rely on the knowledge of the \emph{current} state of the system, i.e. applications that need fresh information at the destination. Under heavy loads, the LCFS queue is frequently replacing its HoL packet with fresher packets. We expect that the higher the packet generation rate at the sources, the lower the average age at the destination, regardless of the queue backlog. LCFS queues are not commonly found in communication systems. Not surprisingly, LCFS is not one of the queueing discipline (qdisc) options in Linux nor in the Software Defined Radios (SDRs) we utilized for implementing WiFresh. In both cases, the standard queueing discipline is FCFS. 

\begin{figure}[t]
\begin{center}
\resizebox{.7\columnwidth}{!}{\includegraphics{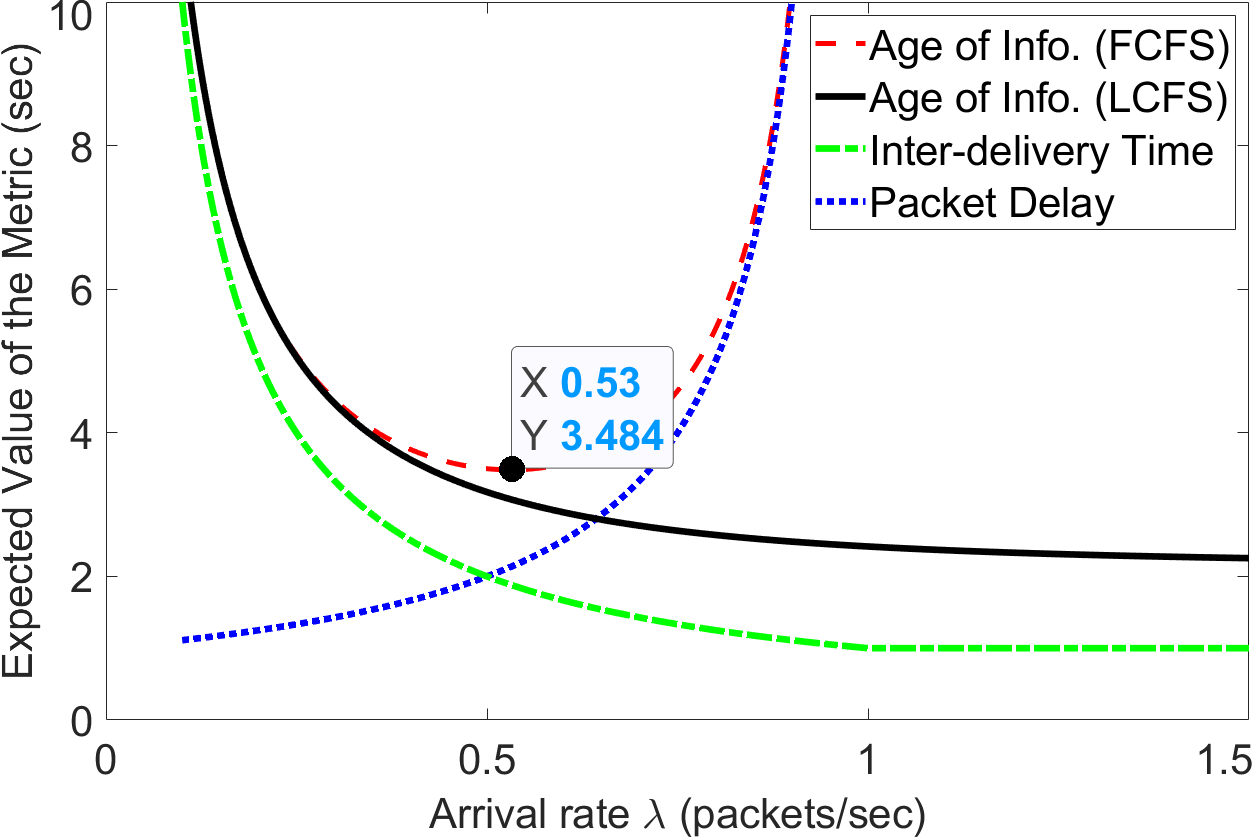}}
\end{center}
\vspace{-0.2cm}
\caption{\label{fig5.AoI_tradeoff_plot} Expected delay, interdelivery time and average age of an M/M/1 queueing system with service rate of $\mu=1$.} 
\vspace{-0.2cm}
\end{figure}

\textbf{Comparing FCFS and LCFS.} Consider an M/M/1 queueing system with infinite queue size, fixed packet service rate of $\mu=1$ packet per second and variable packet generation rate $\lambda$, employing either FCFS or LCFS discipline. In Fig.~\ref{fig5.AoI_tradeoff_plot}, we display the time-average expected age for FCFS and LCFS, the expected packet delay and the expected interdelivery time. The analytical expressions for the AoI associated with FCFS and LCFS queues were obtained in \cite{AoI_update} and \cite{AoI_management}, respectively, and the expressions for packet delay and interdelivery time can be found in \cite{Mor_book}. 

\textbf{Choice of LCFS for WiFresh.} From Fig.~\ref{fig5.AoI_tradeoff_plot}, we can see that the minimum time-average age for FCFS queues is achieved at moderate loads, in particular $\lambda/\mu\approx 0.53$, while for LCFS queues the higher the packet generation rate $\lambda$, the lower the average age. In addition, LCFS outperforms FCFS for every packet generation rate $\lambda$. As discussed in \textsection\ref{sec5.related}, LCFS was shown to be the optimal queueing discipline in different settings including single queue systems \cite{AoI_discipline,AoI_management,AoI_LIFO}, single-hop wireless networks \cite{AoI_LGFS19_1} and multi-hop wireless networks \cite{AoI_LGFS19_2}. Thus, we propose to use the LCFS discipline in WiFresh.

\textbf{Effect of dropping packets.} LCFS queues transmit the freshest packet first. Notice that when a packet with \emph{older information} is delivered to the destination after a packet with \emph{fresher information}, the freshness of the information is not affected and, thus, the value of $\AoI_i(t)$ \emph{remains unchanged}. Hence, if packets with older information were dropped at the source as soon as a fresher packet arrived to the LCFS queue, the information freshness at the destination and the evolution of $\AoI_i(t)$ over time would not be affected. It follows that, \emph{from the perspective of AoI, a LCFS queue is equivalent to a head-drop FCFS queue of size $1$ packet, in which only the freshest packet is kept}. The advantage of dropping older packets at the source is saving communication resources. One possible disadvantage is that dropped packets might contain useful information. For example, in a position tracking system, older packets can be used to predict future movement. 

\subsection{Multiple Access Mechanism}\label{sec5.MAC}
\begin{figure}[t]
\begin{center}
\resizebox{.8\columnwidth}{!}{\includegraphics{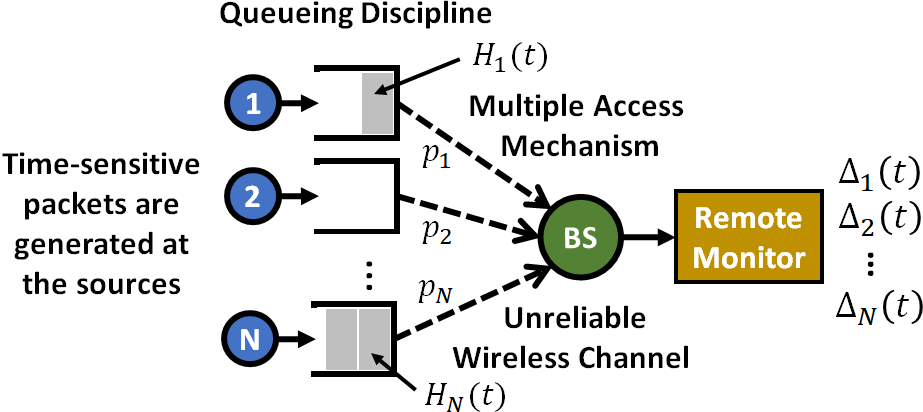}}
\end{center}
\vspace{-0.2cm}
\caption{\label{fig5.communication_network} Illustration of the wireless network.}
\vspace{-0.2cm}
\end{figure}

Consider the network in Fig.~\ref{fig5.communication_network} with $N$ sources sending time-sensitive information to the remote monitor via the wireless BS. Packets are generated at the sources and enqueued in separate queues. The multiple access mechanism controls the method utilized by each of the $N$ sources for sharing the common wireless channel. In this section, we compare two types of multiple access mechanism, Random Access and Polling, in terms of information freshness.

To capture the freshness of the information \emph{in the network}, we define the \emph{expected network AoI (NAoI)} as
\begin{equation}\label{eq5.EWSAoI}
\textstyle\lim_{T\rightarrow\infty}\frac{1}{TN}\int_{t=0}^T \sum_{i=1}^N \mathbb{E}\left[\AoI_{i}(t)\right]dt \; ,
\end{equation}
where $T>0$ is the time-horizon. To minimize NAoI, the multiple access mechanism should attempt to: i) provide high communication efficiency by preventing packet collisions, minimizing the effects of external interference, and reducing control overhead; and ii) prioritize transmissions from sources with high current age $\AoI_i(t)$ and favorable channel conditions. 

\textbf{Random Access} is a widely deployed class of multiple access mechanisms, e.g. WiFi, ZigBee, Wireless Body Area networks \cite{Slotted_ALOHA_use}, and traditional cellular systems such as GSM. The fundamental idea is that, when a source has a packet to transmit, it uses a randomized algorithm to contend for channel access. Randomization is employed to reduce the probability of two or more sources transmitting packets simultaneously, which would result in a packet collision. Some advantages of Random Access are simplicity, decentralization and low control signaling overhead. Some disadvantages are the probability of packet collision that increases with the number of sources $N$, the susceptibility to external interference, and the distributed operation that makes it challenging to implement a dynamic transmission prioritization based on parameters such as age $\AoI_i(t)$ and/or current channel conditions. 

\textbf{Polling mechanism} is a well-known alternative to Random Access \cite{Gil_WiFi}. The BS coordinates the communication in the network by sending \emph{poll packets} to the sources selected for transmission. 
The BS selects the next source to poll based on the \emph{scheduling policy}, which may be a function of dynamic parameters such as age $\AoI_i(t)$ and/or current channel conditions. 
The polling mechanism attempts to leverage all the available communication resources, and it does not back-off when transmission errors occur due to the unreliability of the wireless channel or due to external interference. It is the role of the scheduling policy to estimate the channel conditions and adapt future scheduling decisions accordingly. The polling mechanism attempts to maximize communication efficiency and enables dynamic prioritization, making it suitable for large-scale time-critical applications. 


Two important challenges associated with polling mechanisms are the control overhead and the choice of scheduling policy. 
\emph{Control overhead:} the BS transmits a poll packet before receiving each data packet. In contrast, Random Access may require that the BS transmit an acknowledgment packet following the reception of each data packet. Hence, the control overhead of both mechanisms is comparable. 
\emph{Scheduling policy:} the BS dynamically chooses the next source to poll. Evidently, a naive policy can degrade the performance. Next, we discuss scheduling policies that are designed to optimize the information freshness in the network. 


\subsection{Scheduling Policy}\label{sec5.Scheduling}
The problem of obtaining an optimal scheduling policy for single-hop wireless networks in terms of information freshness was shown in \cite{AoI_scheduling} to be NP-hard. Numerous heuristic policies based on Approximate Dynamic Programming \cite{YuPin17}, Restless Multi-Armed Bandits \cite{igorTON18,Book19} and Lyapunov Optimization \cite{igorTON18,igorTON18_2,igorMobiHoc,AoI_sync,Book19} have been proposed in the literature. \emph{This paper is the first to implement an AoI-based scheduling policy in a real network.} The Max-Weight policy is chosen for WiFresh because it is intuitive, low-complexity and has superior performance \cite{igorMobiHoc}. 

\textbf{Max-Weight (MW) policy.} Consider the network in Fig.~\ref{fig5.communication_network} employing LCFS queues and a Polling mechanism. Assume that $t$ is the current decision time of the next poll packet. Let $p_i\in(0,1]$ be the channel reliability associated with source $i$, namely the probability of a successful reception of a data packet following the transmission of a poll packet to source $i$. Let $\tau^{HoL}(t)$ be the time-stamp of the current Head-of-Line packet from source $i$ at time $t$ and let $\sysTime_i(t):=t-\tau^{HoL}(t)$ be the \emph{current system time of this HoL packet}. Notice that if this HoL packet were delivered to the BS at time $t$, then $\sysTime_i(t)$ would be the associated packet delay and the age would be reduced from $\AoI_i(t)$ to $\sysTime_i(t)$, as illustrated in Fig.~\ref{fig5.AoI_illustration}. Hence, the difference $\AoI_i(t)-\sysTime_i(t)$ represents the \emph{potential age reduction} of polling source $i$ at time $t$. Assume that the scheduling policy knows $\AoI_i(t)$, $\sysTime_i(t)$ and $p_i$, and denote $\mathcal{I}(i,t):=p_i(\AoI_i(t)-\sysTime_i(t))^2$ as the index of source $i$ at time $t$. Then, the MW policy selects, at every decision time $t$, the source $i^*(t)$ with highest value of $\mathcal{I}(i,t)$, with ties being broken arbitrarily. 
Intuitively, the MW policy is polling the source with highest \emph{weighted potential age reduction}. The MW policy was developed in \cite{igorMobiHoc} where we also obtained performance guarantees in terms of AoI. 
To implement the MW policy in a real network, we augment WiFresh with algorithms that estimate $\AoI_i(t)$, $\sysTime_i(t)$ and $p_i$ over time, as described in \textsection\ref{sec5.WiFreshRT}. 


\section{Design and Implementation}\label{sec5.design}
In this section, we discuss the design and implementation of 
WiFresh Real-Time and WiFresh App. 
Prior to delving into the details, we describe the main challenges.

\subsection{Challenges}\label{sec5.challenges}
\textbf{Complexity of implementation.} To achieve high performance, the LCFS queues, the Polling mechanism, and the MW policy were \emph{fully implemented in FPGAs} with 10 MHz clocks, enabling WiFresh Real-Time to make the scheduling decision and trigger the transmission of the next poll packet in approximately $20$ microseconds. Keeping this time-interval short and limiting the length of the poll packet are important factors in reducing the control overhead and achieving high performance. 
The main challenge of implementing WiFresh Real-Time at the \emph{MAC layer} is the complexity associated with implementing numerous real-time functions using hardware-level programming.

\textbf{Barrier to adoption.} Targeting an alternative implementation of WiFresh that could be easily integrated into applications that already run over WiFi such as \cite{Kiva,KivaWiFi,Alibaba,PerformanceDSRC,SimVsRealDSRC,VehicleToInfrastructureDSRC,AppLevelDSRC,Drones,DistributedRobotFormation,RobotFormation,ASTRO,DroneCinema,DOOR-SLAM,CooperativeSLAM,MultiRobotSlam,DARPASubTDrones,MultiRobotMapping,FarmBeats,FarmBeats2}, we propose WiFresh App which is implemented in Python~$3$ and runs at the \emph{Application layer}, without modifications to lower layers of the networking protocol stack. The main challenge of WiFresh App is in the design of a Python application that is capable of driving a standard WiFi network (with FCFS queues and Random Access) to behave as WiFresh (with LCFS queues and Polling mechanism with MW policy). This design is discussed in \textsection\ref{sec5.WiFreshApp}. 

\textbf{Bridging theory and practice.} 
Theoretical works on AoI often assume that: 1) nodes are synchronized; 2) nodes generate packets on-demand or according to known stochastic processes; 3) each node is associated with a single type of information such as position, inertial measurements or images; 4) each data packet contains a complete information update; 5) channel reliabilities $\{p_i\}_{i=1}^N$ are fixed and known; and/or 6) system times of HoL packets $\{\sysTime_i(t)\}_{i=1}^N$ are known. 
To leverage the theory and implement an AoI-based network architecture, 
we augment WiFresh with algorithms that synchronize clocks, dynamically learn $\{p_i\}_{i=1}^N$ and $\{\sysTime_i(t)\}_{i=1}^N$, manage sources with multiple information types, and manage packet fragmentation.

\textbf{Fragmentation of information updates.} The age is reduced when fresh information is received at the destination. The evolution of $\AoI_i(t)$, as described in \textsection\ref{sec5.background}, assumes that each data packet contains a complete information update. To accommodate large information updates, such as images, WiFresh has to manage packet fragmentation. Two issues are discussed below.

The first issue is when to reduce the age $\AoI_i(t)$. In general, the age $\AoI_i(t)$ can be reduced (or partially reduced) upon reception of a subset of fragments. In this work, fragmentation is used for transmitting images and, in this case, it makes sense to reduce age only when all fragments are received. The second issue is whether the LCFS queue should replace the HoL packet as soon as a new information update arrives, or if the LCFS queue should wait until all fragments from the previous information update are delivered before replacing the HoL packet. Notice that if information updates are generated with a high rate, then replacing the HoL packet as soon as a new information update arrives may hinder the \emph{complete transmission} of information updates. For this reason, in this work, we choose to transmit all fragments before replacing the information update at the LCFS queue. 

WiFresh Real-Time runs at the MAC layer. Hence, it is blind to the concept of \emph{information} and can only see individual \emph{data packets}. This makes the process discussed above challenging to implement. To overcome this problem, WiFresh Real-Time could gather information regarding fragmentation from other layers of the communication system. In this work, we implement fragmentation only in WiFresh App, which runs at the Application layer and is aware of information updates. Recall that information updates are generated and received by the Application layer. 

\subsection{Design of WiFresh Real-Time}\label{sec5.WiFreshRT}
\begin{figure}[t]
\begin{center}
\resizebox{0.85\columnwidth}{!}{\includegraphics{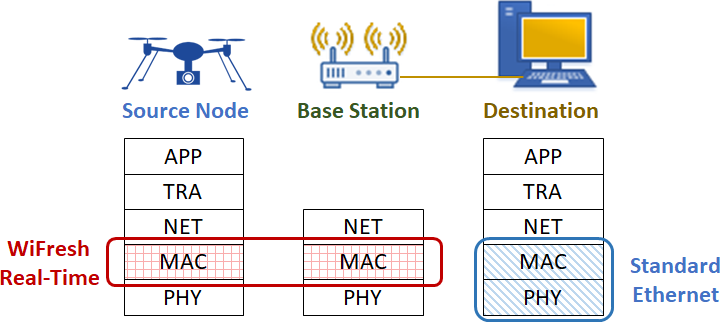}}
\end{center}
\vspace{-0.2cm}
\caption{\label{fig5.WiFreshRTDes} Layers of the WiFresh RT system.} 
\vspace{-0.2cm}
\end{figure}

In this section, we describe WiFresh Real-Time (WiFresh RT) in detail. 
WiFresh RT is implemented at the \emph{MAC layer}, as illustrated in Fig.~\ref{fig5.WiFreshRTDes}. 
Next, we describe the main functions associated with the sources and the base station. 


\textbf{WiFresh RT source.} The source generates information updates in the Application layer and forwards them to lower layers of the networking protocol stack. When a data packet arrives at the MAC layer, WiFresh RT appends a time-stamp to the packet and then stores it in a head-drop FCFS queue of size $1$ packet. Recall that this queue keeps only the freshest packet and discards older packets. The source can be in one of two states: 1) waiting for a poll packet from the BS; or 2) transmitting the freshest data packet to the BS. 
Upon receiving a poll packet, 
if the queue is empty, the source transmits an \emph{empty packet} to the BS. 
The empty packet is used by the BS to differentiate between not receiving data due to a transmission error or due to an empty queue at the source, which impacts the estimation of $p_i$ and $\sysTime_i(t)$ at the BS. After transmitting either the data packet or the empty packet, the source goes back to waiting for the next poll packet.

\textbf{WiFresh RT Base Station.} The BS does not generate data packets. Its main responsibility is to coordinate the communication in the network. The BS can be in one of two states: 1) waiting for a data packet; or 2) transmitting a poll packet. While waiting for a data packet, the BS keeps track of the waiting period. If the waiting period exceeds $100$ microseconds or a data packet is received, the BS updates its estimate of the network state $(\hat{\AoI}_i(t),\hat{\sysTime}_i(t),\hat{p}_i(t))_{i=1}^N$, where $\hat{\AoI}_i(t),\hat{\sysTime}_i(t)$ and $\hat{p}_i(t)$ are the estimates of $\AoI_i(t),\sysTime_i(t)$ and $p_i$, respectively, at time $t$. These estimates are used by the MW policy to calculate $i^*(t)=\argmax{\{\;\hat{p}_i(t)(\hat{\AoI}_i(t)-\hat{\sysTime}_i(t))^2\}}$. After transmitting a poll packet to source $i^*(t)$, the BS goes back to waiting for the next data packet.  
The algorithms used to estimate $\AoI_i(t)$, $p_i$ and $\sysTime_i(t)$ are discussed next.

\textbf{Clock synchronization} is needed to accurately compute $\AoI_i(t):=t-\tau_i(t)$, where $t$ is the current time \emph{measured by the BS} and $\tau_i(t)$ is a time-stamp \emph{created by source} $i$. If clocks are not synchronized, the values of $\AoI_i(t)$ for different sources may have different biases, which may lead to poor scheduling decisions by the MW policy. To estimate the time-stamp offset between each source and the BS, and obtain the estimates $\{\hat{\AoI}_i(t)\}_{i=1}^N$, some possible approaches are: adding GPS antennas to every source in the system and then using GPS time; synchronizing the Operating System (OS) of every source using the Network Time Protocol (NTP) \cite{NTP} via the Internet and then using the OS time; or implementing a synchronization algorithm as part of WiFresh. In WiFresh RT we use the OS time. In WiFresh App we implement a built-in synchronization algorithm based on NTP.

\textbf{Learning channel reliability.} To estimate the value of $p_i\in(0,1]$ associated with each source $i$, we implement a low-complexity estimator.  Let $\mathcal{P}_i(t)$ be the number of poll packets transmitted to source $i$ in the last $\mathcal{W}$ seconds and let $\mathcal{D}_i(t)$ be the number of \emph{data packets} and \emph{empty packets} successfully received from source $i$ in the same period. Then, the estimate of $p_i$ at time $t$ is given by $\hat{p}_i(t)=(\mathcal{D}_i(t)+1)/(\mathcal{P}_i(t)+1)$. 
We choose a time window of $\mathcal{W}=0.5$ seconds. Notice that when the number of poll packets $\mathcal{P}_i(t)$ is low, the estimate $\hat{p}_i(t)$ tends to be optimistic, i.e. higher. In particular, when $\mathcal{P}_i(t)=\mathcal{D}_i(t)=0$, we have $\hat{p}_i(t)=1$. This high value of $\hat{p}_i(t)$ when the number of poll packets is low creates an incentive for the MW policy to select sources that have not been polled recently. 

To determine the changes in $\mathcal{D}_i(t)$ and $\mathcal{P}_i(t)$ for each source $i$ over time, we log the transmission and reception events within the window $\mathcal{W}$ using arrays. This log is created at the on-board processor of the SDR (as opposed to the FPGA) in order to spare the limited FPGA resources. 
The disadvantage of keeping the log at the processor is the added round-trip communication delay between on-board processor and FPGA which is of approximately $500$ microseconds. Since $\hat{p}_i(t)$ represents the average channel reliability in the last $\mathcal{W}=0.5$ seconds, it follows that this relatively small round-trip communication delay has a negligible impact on the performance of the MW policy. The estimate of $p_i$ is the only portion of WiFresh RT which is not fully implemented at the FPGA.

\textbf{Learning the system times.} Recall that the difference $\AoI_i(t)-\sysTime_i(t)$ represents the \emph{potential age reduction} of polling source $i$ at time $t$ and that the MW policy wishes to use this difference for selecting the appropriate source to poll. The problem is that the MW policy does not know the system times of the HoL packets $\{\sysTime_i(t)\}_{i=1}^N$, which are only known by the respective sources, as illustrated in Fig.~\ref{fig5.communication_network}. One approach for estimating $\sysTime_i(t)$ could be to develop an algorithm that generates estimates $\hat{\sysTime}_i(t)$ based on the entire history of transmission and reception events, especially the sequence of previously received time-stamps. The main drawback of this approach is its computational complexity. A less accurate but much simpler approach is to estimate $\sysTime_i(t)$ based on the latest received packet only. In particular, we know that when the freshest data packet from source $i$ is received at time $t$, the potential age reduction of polling source $i$ again at time $t$ is (most likely\footnote{The potential age reduction of polling source $i$ again at time $t$ might be greater than zero if source $i$ generates a new data packet while the previous data packet was being transmitted. We assume that this is an unlikely event and neglect its effect.}) zero, which is represented by $\sysTime_i(t)=\AoI_i(t)$. Similarly, when an empty packet is received at time $t$, the potential age reduction of polling source $i$ again at time $t$ is (most likely) zero. Hence, we can estimate $\sysTime_i(t)$ using the following mechanism: 
\begin{itemize}
\item $\hat{\sysTime}_i(t)\leftarrow \hat{\AoI}_i(t)$ following the successful reception of a data packet or an empty packet from source $i$ at time $t$; and 
\item $\hat{\sysTime}_i(t)$ remains constant over time while no packet is received. 
\end{itemize}
This low-complexity mechanism for obtaining $\hat{\sysTime}_i(t)$ prevents the MW policy to repeatedly schedule the same source~$i$ with a high age $\hat{\AoI}_i(t)$ and an \emph{empty queue}. In \textsection\ref{sec5.Network_Scalability}, we compare the MW policy with the Maximum Age First (MAF) policy which schedules the source $i$ with highest current age $\hat{\AoI}_i(t)$, disregarding the estimates $\hat{p}_i(t)$ and $\hat{\sysTime}_i(t)$. We show that, as expected, MW outperforms MAF in every experiment. 


Estimation errors in 
$\hat{\AoI}_i(t)$, $\hat{p}_i(t)$ and/or $\hat{\sysTime}_i(t)$ affect the performance of WiFresh RT only when they result in poor scheduling decisions. 
In \textsection\ref{sec5.evaluation}, we evaluate the performance of WiFresh RT using the low-complexity mechanisms described in this section and show that WiFresh RT achieves low NAoI in a variety of network settings. 
%
%
%
Next, we discuss the implementation of WiFresh RT.


\subsection{Implementation of WiFresh Real-Time}\label{sec5.WiFreshRTImp}
We implement WiFresh RT in the SDR testbed in Fig.~\ref{fig5.WiFreshRTtestbed} composed of one NI USRP 2974 operating as the wireless base station, and ten sources: seven NI USRP 2974 and three NI USRP 2953R. The code is developed using 
a modifiable WiFi reference design \cite{80211AFW} with Transport layer based on UDP, MAC layer based on the Distributed Coordination Function (DCF), PHY layer based on the IEEE 802.11n standard with center frequency $2.437$ GHz, $20$ MHz bandwidth and a fixed MCS index of $5$. We use this WiFi reference design as a starting point to implement WiFresh RT. 


The WiFi reference design is composed of two parts: the Host code (running at the on-board Intel i7 6822EQ 2 GHz Quad Core processor) and the FPGA code (running at the Xilinx Kintex-7 XC7K410T FPGA). 
The Host code is responsible for the generation of data packets, radio configuration, and displaying measurements and plots. The FPGA code is responsible for processing data packets, generating control packets (e.g. Clear-to-Send, Request-to-Send and Acknowledgments), accessing the wireless channel using DCF, time management (e.g. Interframe spaces and timeouts), etc. 
The FPGA code allows us 
to implement real-time functions at the hardware level. The FPGA clock is of 10 MHz, meaning that these functions run at the microsecond time-scale. 
For implementing WiFresh RT, we developed several new real-time functions at the FPGA, including:
1) Polling mechanism; 2) Max-Weight policy; 3) head-drop FCFS queue with size 1 packet; 4) time-stamp processing; 5) learning algorithms; and 6) measurement logs.

\subsection{Design of WiFresh App}\label{sec5.WiFreshApp}
\begin{figure}[t]
\begin{center}
\resizebox{.8\columnwidth}{!}{\includegraphics{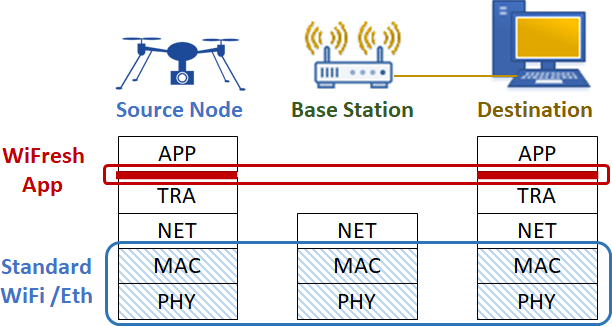}}
\end{center}
\vspace{-0.2cm}
\caption{\label{fig5.WiFreshAppDes} Layers of the WiFresh App system.} 
\vspace{-0.2cm}
\end{figure}

WiFresh App is an implementation of WiFresh that aims to be easily integrated into time-sensitive applications that already run over WiFi such as \cite{Kiva,KivaWiFi,Alibaba,PerformanceDSRC,SimVsRealDSRC,VehicleToInfrastructureDSRC,AppLevelDSRC,Drones,DistributedRobotFormation,RobotFormation,ASTRO,DroneCinema,DOOR-SLAM,CooperativeSLAM,MultiRobotSlam,DARPASubTDrones,MultiRobotMapping,FarmBeats,FarmBeats2}. WiFresh App is implemented 
in Python and runs at the Application layer, without modifications to lower layers of the networking protocol stack, as illustrated in Fig.~\ref{fig5.WiFreshAppDes}. It is designed to drive a standard WiFi network (with FCFS queues and Random Access) to behave as WiFresh (with LCFS queues and Polling mechanism with MW policy). WiFresh App contains all elements of WiFresh RT and some additional features, namely fragmentation of large information updates, a simple built-in synchronization algorithm, and support for sources that generate multiple types of information. Next, we describe WiFresh App.

\textbf{WiFresh App source.} The source generates information updates at the Application layer. WiFresh App time-stamps the information updates and stores them in a LCFS queue, which is implemented using a \emph{Python LIFO stack}. The LCFS queue releases a single information update only when the source receives a poll packet from the destination. If the released information update fits into a single data packet, this information update is encapsulated into a data packet and forwarded to lower layers of the networking protocol stack. Otherwise, the information update is fragmented, stored and the first data packet is forwarded. Fragments are stored in a FCFS queue which is separate from the LCFS queue containing information updates. Upon receiving the next poll packet acknowledging the first fragment, the source forwards the second fragment, and so on, until all fragments are successfully delivered to the destination. When the poll packet acknowledging the final fragment is received, the LCFS queue releases the next information update, which is then fragmented, stored and transmitted following the same procedure. 

When a fragment reaches the source's MAC layer, WiFi stores it in a FCFS queue and transmits it to the destination using Random Access. Ideally, since the destination only generates a new poll packet after the previous fragment is received, there should be \emph{at most one source} attempting transmission using Random Access at any given time. This means that, even when all sources are generating information updates with a high rate, the underlying WiFi network is handling one data packet at a time, leading to low packet delay and low probability of packet collision. 
When transmission errors occur, WiFi may attempt to retransmit the data packet. Notice that by implementing LCFS queues and Polling mechanisms with MW policy at the Application layer, \emph{WiFresh App is driving the underlying WiFi network to behave as a WiFresh network.}

\textbf{WiFresh App destination.} Similarly to the WiFresh RT Base Station, the destination in WiFresh App generates poll packets, implements a timeout of $300$ milliseconds, updates its estimate of the network state $(\hat{\AoI}_i(t),\hat{\sysTime}_i(t),\hat{p}_i(t))_{i=1}^N$, and uses the MW policy to decide which source to poll next. The main differences are that: 1) the scheduling decisions are made at the Application layer at the millisecond time-scale, as opposed to the MAC layer at the microsecond time-scale; and 2) the destination manages the fragmentation procedure described above, uses a built-in clock synchronization algorithm based on the \emph{on-wire protocol} that is part of NTP \cite[\textsection8]{NTP} to estimate the current age $\hat{\AoI}_i(t)$, and supports sources that generate multiple types of information.


\textbf{Multiple information types per source.} The age is associated with a single type of information such as position, inertial measurements or images. In a network with sources that generate multiple types of information, we create for each tuple (source, information type) a separate instance of WiFresh App with independent LCFS queue, 
age $\hat{\AoI}_i(t)$, channel reliability $\hat{p}_i(t)$, and system time $\hat{\sysTime}_i(t)$. 
The destination treats each instance of WiFresh App at the sources as an independent entity, and sends individual poll packets to each of them. 

\subsection{Implementation of WiFresh App}\label{sec5.WiFreshAppImp}
We implement WiFresh App in a Raspberry Pi (Raspi) testbed composed of one desktop computer operating as the destination, one Raspberry Pi 3B+ with a WiFi USB adapter operating as the wireless BS, and twenty four sources: ten Raspberry Pi 3B+ fetching data from sensors and fourteen Raspberry Pi Zero W generating synthetic data that emulates the sensors. For the measurements in \textsection\ref{sec5.evaluation}, sources are static and placed indoors. The distance between sources and destination is between 2 and 3 meters. In Fig.~\ref{fig5.WiFreshAppTestbed}, we display some of the sources\footnote{The remote control cars and battery packs are used for running tests outdoors. The measurements discussed in this paper were performed indoors.} and the three sensors described below:
\begin{itemize}
\item cameras (Arducam 5 Megapixels 1080p) generating jpg images with resolution 256x144 pixels and size of approximately $19$ kbytes at a rate of $2$ Hz.
\item Inertial Measurement Units (Pololu MinIMU-9 v5 Gyro, Accelerometer, and Compass) generating information updates of size $20$ bytes at a rate of $100$ Hz; and
\item GPS units (Stratux GPYes 2.0 u-blox 8) generating information updates of size $50$ bytes at a rate of $1$ Hz;
\end{itemize}
To create synthetic GPS data for indoor environments, we use a NMEA sentence generator \cite{GPSsynthetic} that emulates the GPS unit.


The Raspberry Pis run the Raspbian Stretch OS and communicate via WiFi, in particular the IEEE 802.11g standard at $2.4$ GHz. WiFresh App is implemented using Python 3. 
The main functionalities we developed are: 1) Polling mechanism; 2) Max-Weight policy; 3) LCFS queue; 4) fragmentation management; 5) time-stamp processing; 6) learning algorithms; 7) interface with sensors; 8) synthetic generation of data packets emulating each type of sensor; 9) graphical user interfaces; and 10) measurement logs. The Transport, Network, MAC and PHY layers were kept unchanged, as illustrated in Fig.~\ref{fig5.WiFreshAppDes}. \emph{WiFresh App is built over standard UDP, IP and WiFi protocols.}

\section{Experimental Results}\label{sec5.evaluation}
In this section, we evaluate the performance of WiFresh in a dynamic indoor office space with \emph{multiple external sources of interference} such as mobile phones, laptops and campus WiFi base stations. We evaluate WiFresh RT and WiFresh App, and compare them with other communication systems. In particular, using the SDR testbed described in \textsection\ref{sec5.WiFreshRTImp}, we compare:
\begin{itemize}
\item \textbf{WiFresh RT:} as described in \textsection\ref{sec5.WiFreshRT};
\item \textbf{WiFresh RT FCFS:} identical to WiFresh RT but with sources employing FCFS queues;
\item \textbf{WiFi UDP FCFS}: UDP over standard WiFi; and
\item \textbf{WiFi UDP LCFS}: UDP over standard WiFi but with sources employing LCFS queues instead of FCFS queues.
\end{itemize}
In addition, using the Raspi testbed described in \textsection\ref{sec5.WiFreshAppImp}, we compare:
\begin{itemize}
\item \textbf{WiFresh App:} as described in \textsection\ref{sec5.WiFreshApp};
\item \textbf{WiFresh Max. Age First (MAF)}: identical to WiFresh App but with a scheduling policy that, at every decision time $t$, selects the source $i^*(t)$ with highest current age $\AoI_i(t)$. The MAF policy was proposed and analyzed in \cite{igorTON18};
\item \textbf{WiFi UDP FCFS}: UDP over standard WiFi;
\item \textbf{WiFi TCP FCFS}: TCP over standard WiFi; and
\item \textbf{WiFi ACP FCFS}: Age Control Protocol (ACP) over standard WiFi. ACP is a Transport layer protocol recently proposed in \cite{AoI_ACP} that adapts the packet generation rate of each source $i$ in order to minimize the NAoI in Eq.\eqref{eq5.EWSAoI}. Recall that in our testbed, the packet generation rate is fixed and determined by the associated sensor. Hence, in our implementation of ACP, we approximate the target packet generation rate by regularly discarding some of the packets before they reach the FCFS queue. 
\end{itemize}
Next, we present experimental evaluations of WiFresh. Each experiment runs for $10$ minutes.

\subsection{Single Source with High Load}\label{sec5.Single_Source}
In this section, we consider a network with a destination, a wireless BS and a \emph{single source} generating packets of $150$ bytes with rate $\lambda\in\{5,6,7\}$ kHz. These short packets of $150$ bytes represent status updates, and different values of $\lambda$ represent different levels of congestion. In Tables~\ref{tab.single_source_SDR} and \ref{tab.single_source_Raspi}, we measure the time-average AoI (in seconds) and the effective throughput (in Mbps). The effective throughput is measured at the Application layer of the destination and, thus, it refers to the number of \emph{useful bits} received per second. In Table~\ref{tab.single_source_SDR}, we consider WiFresh RT and WiFi UDP FCFS in the SDR testbed, and in Table~\ref{tab.single_source_Raspi}, we consider WiFresh App and WiFi UDP FCFS in the Raspi testbed. 

\begin{table}[t]
\caption{\label{tab.single_source_SDR} AoI measurements for the SDR testbed. 
}
\vspace{-0.2cm}
\begin{center}
\begin{tabular}{lcccc} 
\toprule
SDR				& \multicolumn{2}{c}{WiFresh RT}& \multicolumn{2}{c}{WiFi UDP FCFS}\\
\midrule
				& AoI 		& Thr. 				& AoI 		& Thr. \\
				&(sec)		&(Mbps)				&(sec)		&(Mbps)\\		
\midrule
$\lambda=5$k 	& 0.003		& 4.866				& 0.306		& 2.406	\\
$\lambda=6$k 	& 0.003		& 4.905				& 0.304		& 2.433	\\
$\lambda=7$k 	& 0.004		& 4.412				& 0.320		& 2.328	\\
\bottomrule
\end{tabular}
\end{center}
\vspace{-0.2cm}
\end{table}

\begin{table}[t]
\caption{\label{tab.single_source_Raspi} AoI measurements for the Raspi testbed.  
}
\vspace{-0.2cm}
\begin{center}
\begin{tabular}{lcccc} 
\toprule
Raspi 			& \multicolumn{2}{c}{WiFresh App}& \multicolumn{2}{c}{WiFi UDP FCFS}\\
\midrule
				& AoI 		& Thr. 				& AoI 		& Thr. \\
				&(sec)		&(Mbps)				&(sec)		&(Mbps)\\		
\midrule
$\lambda=5$k 	& 0.040		& 0.229 			& 224.8 	& 1.242 	\\
$\lambda=6$k 	& 0.046		& 0.197 			& 248.2		& 1.183 	\\
$\lambda=7$k 	& 0.042		& 0.208 			& 242.3		& 1.281 	\\
\bottomrule
\end{tabular}
\end{center}
\vspace{-0.2cm}
\end{table}


\textbf{External interference.} The results in Tables~\ref{tab.single_source_SDR} and \ref{tab.single_source_Raspi} show that when the packet generation rate increases {\color{black}from $5$ kHz to $7$ kHz}, the effective throughput does not change significantly, indicating that {\color{black}sources with $\lambda\geq 5$ kHz are saturated}, i.e. always have data to transmit. 
Table~\ref{tab.single_source_SDR} shows that the throughput of WiFresh RT is higher than the throughput of WiFi UDP FCFS. This is because \emph{WiFresh RT does not back-off} when a transmission error occurs due to the unreliability of the wireless channel or due to collisions with external wireless networks, making WiFresh RT less susceptible to external interference than WiFi UDP FCFS. Recall that WiFresh RT is designed to support large-scale time-critical applications and, to that end, it attempts to maximize its  channel utilization. In contrast, WiFresh App runs over standard WiFi, making it as susceptible to external interference as WiFi UDP FCFS. Table~\ref{tab.single_source_Raspi} shows that the throughput of WiFresh App is lower than the throughput of WiFi UDP FCFS. The main reason for the lower throughput is the \emph{control overhead} associated with running a Polling mechanism over standard WiFi. Notice that acknowledgement packets follow the successful transmission of every poll and data packets, thus increasing the control overhead. Despite the lower throughput, WiFresh App significantly outperforms WiFi UDP FCFS in terms of AoI, as we see next. 

\begin{figure}[t]
\begin{center}
\resizebox{.99\columnwidth}{!}{\includegraphics{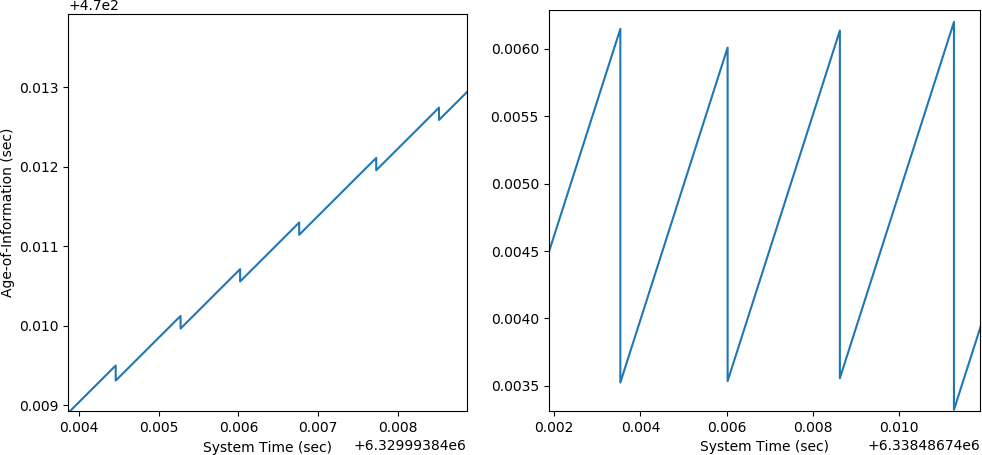}}
\end{center}
\vspace{-0.2cm}
\caption{\label{fig5.AoI_FCFS_LCFS} Age $\AoI_i(t)$ evolution over time in the Raspi testbed with $\lambda=6$ kHz. On the LHS we have WiFi UDP FCFS and on the RHS we have WiFresh App, which uses LCFS.}
\vspace{-0.2cm}
\end{figure}

\textbf{Queueing discipline.} The results in Tables~\ref{tab.single_source_SDR} and \ref{tab.single_source_Raspi} show that WiFresh RT and WiFresh App can improve age by two orders of magnitude when compared to WiFi UDP FCFS. In this single source scenario, the performance gain comes in most part from using LCFS instead of FCFS. In Fig.~\ref{fig5.AoI_FCFS_LCFS}, we compare the age $\AoI_i(t)$ evolution over time for systems employing FCFS and LCFS. Notice that the high packet generation rate $\lambda$ degrades the age $\AoI_i(t)$ performance of the FCFS queue, and improves the age performance of the LCFS queue. 

\textbf{Queue size.} In all three WiFi UDP FCFS experiments in Table~\ref{tab.single_source_Raspi}, the age $\AoI_i(t)$ grows as in Fig.~\ref{fig5.AoI_FCFS_LCFS} throughout the entire experiment, i.e. for $600$ seconds, giving a time-average age of at least $220$ seconds. This result suggests that the FCFS queue of the Raspberry Pi did not overflow, which would have helped stabilizing the age. In contrast, in all three WiFi UDP FCFS experiments in Table~\ref{tab.single_source_SDR}, the FCFS queue\footnote{The transmission queue of the SDR can store one megabyte of data. Notice that for $\lambda=5$kHz we are generating $0.75$ megabyte per second.} overflows in the first few seconds, limiting the age $\AoI_i(t)$ growth and resulting in a time-average age of around $0.3$ seconds. This suggests that smaller FCFS queues result in better age performance. 

\subsection{Network with Increasing Load}\label{sec5.Arrival_Rates}

\begin{figure}[t]
\begin{center}
\resizebox{.99\columnwidth}{!}{\includegraphics{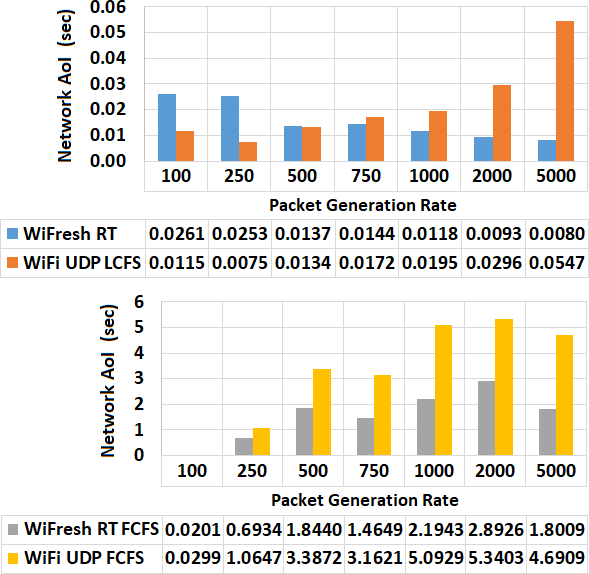}}
\end{center}
\vspace{-0.2cm}
\caption{\label{fig5.plot_rates_USRP} Time-average NAoI measurements for the SDR testbed with ten sources generating {\color{black} packets of $150$ bytes with rate $\lambda\in\{100,250,500,750,1k,2k,5k\}$} Hz.}
\vspace{-0.2cm}
\end{figure}

In this section, we consider a network with a destination, a wireless BS and \emph{ten sources} generating packets of $150$ bytes with rate $\lambda$. 
In Fig.~\ref{fig5.plot_rates_USRP}, we display the NAoI measurements (in seconds) for the SDR testbed employing the following communication systems: 1) WiFresh RT; 2) WiFi UDP LCFS; 3) WiFresh RT FCFS; and 4) WiFi UDP FCFS. 

By comparing the results of WiFresh RT and WiFi UDP FCFS for $\lambda\geq 500$ Hz, we can see that \emph{WiFresh RT improves information freshness by (at least) a factor of} $200$ \emph{when compared to an equivalent standard WiFi network}. To understand how much of this improvement is due to the queueing discipline and how much is due to the multiple access mechanism, we draw additional comparisons. By comparing WiFresh RT and WiFi UDP LCFS, both of which use LCFS queues, we can assess the impact of the multiple access mechanism on NAoI. As expected, the improvement of Polling over Random Access increases with the network congestion. In particular, for $\lambda=5$ kHz, WiFresh RT improves age by a factor of $7$ when compared to WiFi UDP LCFS. To assess the impact of queueing, we compare WiFresh RT and WiFresh RT FCFS, both of which use Polling with MW policy. For $\lambda\geq 500$ Hz, the LCFS queue improves information freshness by (at least) a factor of $100$ when compared to the FCFS queue. \emph{Both the queueing discipline and the multiple access mechanism improve NAoI significantly}, but the effect of queueing is dominant.

\begin{figure}[t]
\begin{center}
\resizebox{.99\columnwidth}{!}{\includegraphics{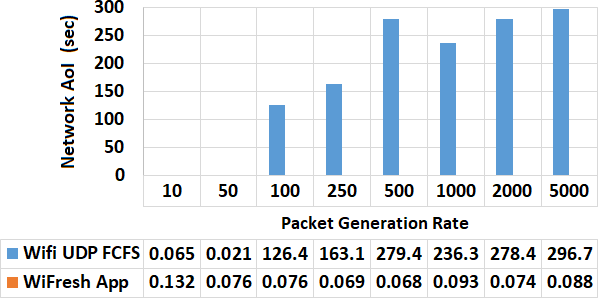}}
\end{center}
\vspace{-0.2cm}
\caption{\label{fig5.plot_rates_Pi} NAoI measurements for the Raspi testbed with ten sources generating packets of $150$ bytes with rate $\lambda\in\{10,50,100,250,500,750,1k,2k,5k\}$ Hz.}
\vspace{-0.2cm}
\end{figure}

In Fig.~\ref{fig5.plot_rates_Pi}, we display the expected NAoI measurements (in seconds) for the Raspi testbed employing the following communication systems: 1) WiFresh App; and 2) WiFi UDP FCFS. 
The results in Fig.~\ref{fig5.plot_rates_Pi} show that for $\lambda\geq 100$ Hz, \emph{WiFresh App improves information freshness by three orders of magnitude when compared to an equivalent standard WiFi network}. We note that WiFi UDP FCFS performs differently in the Raspi and SDR testbeds due to differences in the platforms, and in particular due to \emph{differences in the FCFS queue sizes}. The large FCFS queues at the Raspberry Pis have a negative effect on WiFi UDP FCFS, which amplifies the performance gain of WiFresh App at high packet generation rates $\lambda$.

\textbf{No need for congestion control.} The results in Figs.~\ref{fig5.plot_rates_USRP} and \ref{fig5.plot_rates_Pi} show that the combination of LCFS queues and Polling mechanism with MW policy is the \emph{only in which a higher rate $\lambda$ leads to a lower NAoI}, meaning that the WiFresh architecture eliminates the need for controlling the packet generation rate at
the sources. Notice that any of the other three architectures, which employ either FCFS queues or Random Access, need to control $\lambda$ in order to minimize NAoI.

\subsection{Network with Increasing Size}\label{sec5.Network_Scalability}


In this section, we consider a network with a destination, a wireless BS and $N$ sources, each source generates up to three types of information updates: position information of $50$ bytes at $1$ Hz, inertial measurements of $20$ bytes at $100$ Hz, and images of $19$ kbytes at $2$ Hz. 
Notice that a network with $N$ physical sources can have up to $3N$ sources of information, each source of information with its own independent instance of WiFresh App. 

\begin{figure}[t]
\begin{center}
\resizebox{.99\columnwidth}{!}{\includegraphics{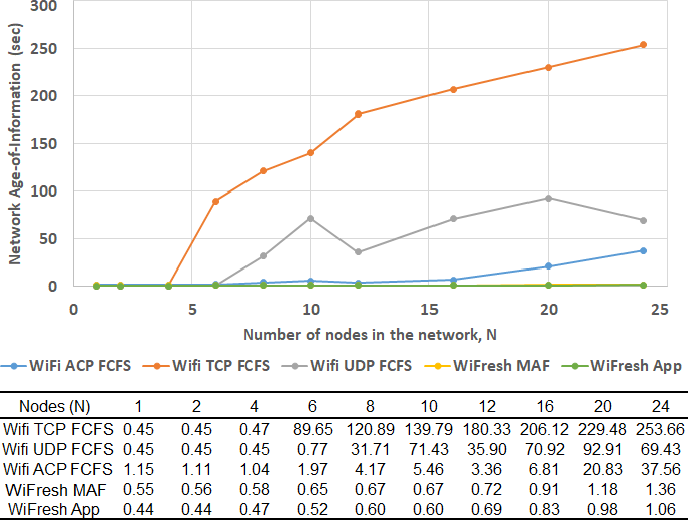}}
\end{center}
\vspace{-0.2cm}
\caption{\label{fig5.table_figure_camera} NAoI measurements for the Raspberry Pi testbed with $N\in\{1,2,4,6,8,10,12,16,20,24\}$ sources generating position information and images.}
\vspace{-0.2cm}
\end{figure}

\begin{figure}[t]
\begin{center}
\resizebox{.99\columnwidth}{!}{\includegraphics{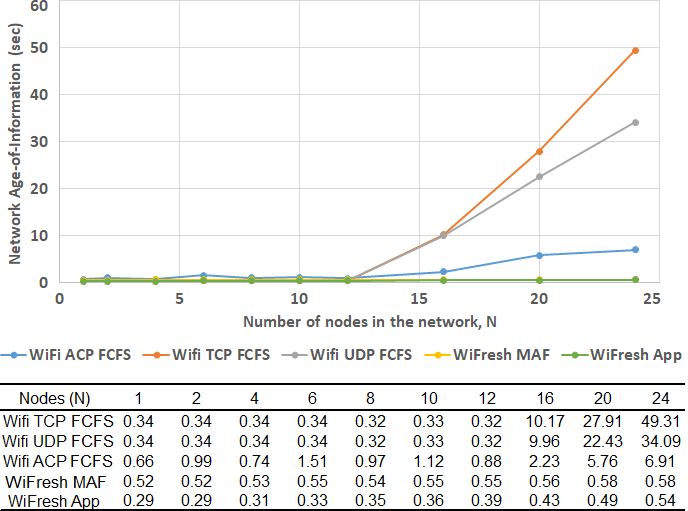}}
\end{center}
\vspace{-0.2cm}
\caption{\label{fig5.table_figure_IMU} NAoI measurements for the Raspberry Pi testbed with $N\in\{1,2,4,6,8,10,12,16,20,24\}$ sources generating position information and inertial measurements.}
\vspace{-0.2cm}
\end{figure}

In Figs.~\ref{fig5.table_figure_camera} and \ref{fig5.table_figure_IMU}, we display the NAoI (in seconds) for the Raspi testbed employing the following communication systems: 1) WiFresh App; 2) WiFresh MAF; 3) WiFi UDP FCFS; 4) WiFi TCP FCFS; and 5) WiFi ACP FCFS. In Fig.~\ref{fig5.table_figure_camera}, we consider sources generating both position information and images, and in Fig.~\ref{fig5.table_figure_IMU}, we consider sources generating both position information and inertial measurements. 

\textbf{TCP over WiFi.} The results in Figs.~\ref{fig5.table_figure_camera} and \ref{fig5.table_figure_IMU} show that WiFi TCP FCFS has the worst performance in terms of AoI. TCP provides reliable and in-order packet delivery by requesting retransmissions and rearranging out-of-order packets before forwarding them to the Application layer. Both of these features can degrade information freshness, especially when sources are generating packets at high rates. 

\textbf{ACP over WiFi.} ACP dynamically adapts the packet generation rates at the sources (by regularly discarding some of the packets) in order to drive the underlying WiFi network to the point of minimum AoI. The results in Figs.~\ref{fig5.table_figure_camera} and \ref{fig5.table_figure_IMU} show that, for $N=20$, WiFi ACP FCFS improves NAoI by a factor of four when compared to WiFi UDP FCFS; in turn WiFresh App improves NAoI by (at least) a factor of forty when compared to WiFi UDP FCFS. 

\textbf{Impact of scheduling policy.} The only difference between WiFresh App and WiFresh MAF is the scheduling policy. MAF schedules the source with highest value of $\hat{\AoI}_i(t)$, and neglects information about channel conditions $\hat{p}_i(t)$ and information about the HoL packet at the source's queue $\hat{\sysTime}_i(t)$. 
For this reason, MAF can often poll sources with poor channel condition or an empty queue, what degrades its NAoI performance. This is a main reason for the performance gap between WiFresh MAF and WiFresh App in Figs.~\ref{fig5.table_figure_camera} and \ref{fig5.table_figure_IMU}.

\textbf{Impact of traffic load.}
The results in Fig.~\ref{fig5.table_figure_camera} show that {\color{black}for $N\geq 16$, WiFresh App improves NAoI by a factor of $65$} when compared to WiFi UDP FCFS, and by a factor of $230$ when compared to WiFi TCP FCFS. The results in Fig.~\ref{fig5.table_figure_IMU} show that {\color{black}for $N\geq 16$, WiFresh App improves information freshness by a factor of $20$} when compared to either WiFi UDP FCFS or WiFi TCP FCFS. The improvement is more evident in Fig.~\ref{fig5.table_figure_camera} since cameras generate more traffic than IMUs. In particular, the camera generates approximately $304$ kbits per second per source while the IMU generates approximately $16$ kbits per second per source. 

\textbf{Summary.} From the measurements in this section, we can see that:
1) the more congested the network, the more prominent is the superiority of WiFresh when compared with WiFi in terms of NAoI;
2) the average NAoI in a WiFresh network scales gracefully with the packet generation rate $\lambda$, as seen in \textsection\ref{sec5.Arrival_Rates}, and with the number of sources $N$, as seen in \textsection\ref{sec5.Network_Scalability}; and
3) WiFresh RT achieves the highest performance in terms of throughput and average NAoI, while WiFresh App achieves high performance and can be easily integrated into time-sensitive applications that already run over WiFi, as discussed in \textsection\ref{sec5.WiFreshApp}.

\section{Final Remarks}\label{sec5.conclusion}

In this paper, we propose WiFresh: an unconventional network architecture that scales gracefully, achieving near optimal information freshness in wireless networks of any size $N$, regardless of the level of congestion $\lambda$, even when the network is overloaded. The superior performance of WiFresh is due to the combination of three elements: Last-Come First-Served queues, Polling Multiple Access mechanism, and Max-Weight scheduling policy. The choice of each of these elements is underpinned by theoretical research. We propose and realize two strategies for implementing WiFresh: WiFresh Real-Time, which is designed to maximize performance, and is implemented at the MAC layer in a network of FPGA-enabled SDRs using hardware-level programming; and WiFresh App which is designed to lower the barriers to adoption, and is implemented at the Application layer, without modifications to lower layers of the networking protocol stack, in a network of Raspberry Pis using Python~3. WiFresh App runs over UDP and standard WiFi, making it easy to integrate into time-sensitive applications that are implemented using WiFi such as \cite{Kiva,KivaWiFi,Alibaba,PerformanceDSRC,SimVsRealDSRC,VehicleToInfrastructureDSRC,AppLevelDSRC,Drones,DistributedRobotFormation,RobotFormation,ASTRO,DroneCinema,DOOR-SLAM,CooperativeSLAM,MultiRobotSlam,DARPASubTDrones,MultiRobotMapping,FarmBeats,FarmBeats2}. Our experimental results show that WiFresh can improve the expected network AoI by two orders of magnitude when compared to an equivalent standard WiFi network. Moreover, our results show that the more congested the network, the more prominent is the superiority of WiFresh when compared with WiFi in terms of information freshness.

\ifCLASSOPTIONcaptionsoff
  \newpage
\fi

\bibliographystyle{IEEEtran}
\bibliography{references}


%

\end{document}